# Quasi-Constant Modulus Design for Nonlinearity-Tolerant Geometric Shaped Four-Dimensional Modulation Format

JUNZHE XIAO, ZEKUN NIU, LYU LI, MINGHUI SHI, WEISHENG HU AND LILIN YI

*State Key Laboratory of Photonics and Communications, Publications Department, School of Integrated Circuit Science and Engineering, Shanghai Jiao Tong University, Shanghai, 200240, China*
*[zekunniu@sjtu.edu.cn](mailto:zekunniu@sjtu.edu.cn)*
*[lilinyi@sjtu.edu.cn](mailto:lilinyi@sjtu.edu.cn)*

**Abstract:** In this paper, the quasi-constant modulus (QCM) property is analyzed and leveraged in the design of nonlinearity-tolerant four-dimensional (4D) modulation formats. Accordingly, we propose a family of QCM-based quadrature amplitude modulation (QCM-QAM) constellations with high spectral efficiencies (SEs) of 9, 11, and 13 bit/4D-sym, respectively. The quasi-constant modulus design theoretically enhances tolerance to fiber nonlinearities. Meanwhile, QCM-QAM is evaluated in an unrepeatered wavelength-division multiplexing (WDM) system over both standard single-mode fiber (SSMF) and non-zero dispersion-shifted fiber (NZDSF). Across all SEs, QCM-QAM demonstrates robust nonlinear tolerance in both SSMF and NZDSF. This is evidenced by a consistent shift of the optimal launch power toward higher values and a significant improvement in effective signal-to-noise ratio (SNR). QCM-QAM also delivers generalized mutual information (GMI) gains of 0.22, 0.09, and 0.21 bit/4D-sym in SSMF, and 0.24, 0.10, and 0.22 bit/4D-sym in NZDSF at the optimal transmitted power, corresponding to the SEs of 9, 11, and 13 bit/4D-sym. Furthermore, QCM-QAM achieves transmission reach extensions of 1.6%, 0.9%, and 1.7% in SSMF, and 1.7%, 1.5%, and 1.8% in NZDSF, respectively, for the three SE levels.

## 1. Introduction

Revolutionary advances in optical transmission technologies have powered the growth of Internet traffic for decades [1]. To satisfying the rising demand in optical fiber communications, enhancing spectral efficiency (SE) of the modulation formats is a crucial way. High-order modulation formats, such as polarization-multiplexed (PM) $M$-ary quadrature amplitude modulation (PM-$M$-QAM), have been widely adopted to enhance SE in wavelength-division multiplexing (WDM) transport systems employing coherent detection [2]. Moreover, constellation shaping techniques such as probabilistic shaping (PS) and geometric shaping (GS) offer additional gains in SE beyond conventional modulation formats [3-5].

Both the PS and the GS can achieve Gaussian capacity distribution to close the shaping gap over the additive white Gaussian noise (AWGN) channel [6]. In PS, symbol probabilities are modified via a range of architectural approaches, which have been extensively studied [7-10]. However, such shaping techniques require a distribution matcher and a distribution dematcher to realize the desired symbol probability distribution. On the other hand, GS achieves shaping gains by optimizing constellation point locations [11-13]. In the optical fiber channel, both the linear shaping gain and the nonlinear shaping gain are achieved by GS simultaneously [14]. Moreover, extending geometric shaping from 2D to multi-dimensional (MD) (e.g., 4D) modulation formats unlocks additional design freedom, yielding further performance gains. [15]. By extending 2D GS to both polarizations, dual-polarization (DP) 4D modulation formats are assumed to better reduce the nonlinear interference (NLI) compared to 2D modulation formats [16].

Prior works [17-25] have aimed to optimize 4D modulation formats for improved performance in the optical fiber channel. The optimization of 4D modulation formats is typically formulated as an optimization problem in 4D space, subject to a transmitted power constraint [21]. The objective of this optimization is to maximize either the generalized mutual information (GMI) or the mutual information (MI). Meanwhile, modern NLI models including enhanced Gaussian noise (EGN) model [26] and their variants [27] are employed in the modulation formats optimization, in order to improve the nonlinear tolerance of the 4D formats in the optical fiber channel [17, 18, 25, 28]. Moreover, constraints such as constant modulus [18, 23], orthant symmetry [17, 24, 25], and shell shaping [13, 29] have been incorporated into constellation optimization to enhance nonlinear tolerance reduce the optimization complexity. In particular, the constant-modulus constraint has been shown to be an effective approach for designing nonlinearity-tolerant 4D modulation formats [18, 23], as it mitigates signal energy fluctuations. However, for high SE modulation formats, enforcing a strict constant-modulus constraint becomes increasingly challenging due to the large number of optimization degrees of freedom (DOFs) [13, 30].

In this paper, the quasi-constant modulus (QCM) property is analyzed and leveraged in the design of nonlinearity-tolerant 4D modulation formats. Accordingly, we propose a family of QCM-based quadrature amplitude (QCM-QAM) modulation formats with high SE of 9, 11, and 13 bit/4D-sym, corresponding to 512-ary, 2048-ary, and 8192-ary 4D constellations, respectively. Unlike the strict constant modulus property in nonlinearity-tolerant modulation formats with relatively SE [18, 23, 31], the energy of 4D symbols in QCM-QAM is only quasi-constant. Nevertheless, the analysis on NLI demonstrates that the quasi-constant modulus property provides comparable mitigation of self-phase modulation (SPM) and cross-phase modulation (XPM). Furthermore, QCM-QAM demonstrates robust nonlinear tolerance across SEs in unrepeatered WDM transmission over standard single mode fiber (SSMF) and non-zero dispersion-shifted fiber (NZDSF), evidenced by a consistent shift of the optimal launch power to higher values and a notable gain in effective signal-to-noise ratio (SNR).

In SSMF, QCM-QAM delivers effective SNR gains of 0.59 dB, 0.58 dB, and 0.54 dB, along with generalized mutual information (GMI) gains of 0.22, 0.09, and 0.21 bit/4D-sym, after transmission over distances of 199 km, 177 km, and 158 km, respectively. In NZDSF, it achieves SNR gains of 0.61 dB, 0.60 dB, and 0.55 dB, and GMI gains of 0.24, 0.10, and 0.22 bit/4D-sym over 183 km, 161 km, and 142 km, respectively. Moreover, QCM-QAM extends transmission reach by 1.6%, 0.9%, and 1.7% in SSMF, and by 1.7%, 1.5%, and 1.8% in NZDSF, for the 512-ary, 2048-ary, and 8192-ary modulation formats, respectively.

The remainder of this paper is organized as follows. Section II reviews the simplified NLI models and shows how both SPM and XPM are mitigated by the 4D constant-modulus property. Section 3 introduces the QCM-QAM family and demonstrates its nonlinearity tolerance with the simplified models. Section IV describes the simulation setup, and Section V presents the performance results achieved by QCM-QAM. Finally, Section VI concludes the paper.

## 2. Simplified nonlinear interference model

In this section, simplified models of SPM and XPM are reviewed, which are employed in the analysis on the NLI relating to the modulation formats.

SPM effect can be approximated expressed in Eq. (3) in [32], in the presence of a moderate amount of chromatic dispersion (CD). By neglecting the polarization cross terms, SPM effect of X and Y polarization can be simplified as:

$$E_x^{spm}(t) \approx E_x(t) * \exp\left\{j\kappa\left[|E_x(t)|^2 + P_{xy}(t)\right] \otimes h_{spm}(t)\right\}, \quad (1)$$

$$E_y^{spm}(t) \approx E_y(t) * \exp\left\{j\kappa\left[|E_y(t)|^2 + P_{xy}(t)\right] \otimes h_{spm}(t)\right\}, \quad (2)$$

where $E_x(t)$ and $E_y(t)$ represents the origin electric fields along the X and Y polarization, and $E_x^{spm}(t)$ and $E_y^{spm}(t)$ denotes electric fields with SPM distortion, $\otimes$ stands for the convo-lution operator, $P_{xy}(t) = |E_x(t)|^2 + |E_y(t)|^2$ is the instantaneous power, $h_{spm}(t)$ is the SPM linear filter derived from the perturbation theory [14], and $\kappa$ is the nonlinear coefficient determined by the fiber parameters.

Likewise, the XPM effect can be simplified by neglecting the polarization cross-terms as [33]:

$$E_x^{xpm}(t) \approx E_x(t) * \exp\left\{j\kappa\left[|E_{x,c}(t)|^2 + P_{xy,c}(t)\right] \otimes h_{xpm,c}(t)\right\}, \tag{3}$$

$$E_y^{xpm}(t) \approx E_y(t) * \exp\left\{j\kappa\left[|E_{y,c}(t)|^2 + P_{xy,c}(t)\right] \otimes h_{xpm,c}(t)\right\}, \tag{4}$$

where and $E_x^{xpm}(t)$ and $E_y^{xpm}(t)$ denotes electric fields with XPM distortion caused by the $c$-th interfering channel. In Eq. (3) and (4), $E_{x,c}(t)$ and $E_{y,c}(t)$ represents the origin electric fields along the X and Y polarization of the $c$-th channel, $P_{xy,c}(t)$ denotes the instantaneous power of the $c$-th channel, and $h_{xpm,c}(t)$ is the XPM linear filter of the $c$-th channel. The XPM filters of different channels have similar expressions, and their effects are cumulative [33]. Without loss of generality, our analysis of XPM focuses on the $c$-th channel and can be extended to other channels.

The expression of the simplified SPM and XPM model can be rewritten as:

$$E_{x/y}^{spm}(t) \approx E_{x/y}(t) * \underbrace{\exp\left\{j\kappa|E_{x/y}(t)|^2 \otimes h_{spm}(t)\right\}}_{\text{Induced by Power Fluctuations in X/Y Pol}} * \underbrace{\exp\left\{j\kappa P_{xy}(t) \otimes h_{spm}(t)\right\}}_{\text{Induced by Fluctuations of Total Power}}, \tag{5}$$

$$E_{x/y}^{xpm}(t) \approx E_{x/y}(t) * \underbrace{\exp\left\{j\kappa|E_{x/y,c}(t)|^2 \otimes h_{xpm,c}(t)\right\}}_{\text{Induced by Power Fluctuations in }c\text{-th Channel X/Y Pol}} * \underbrace{\exp\left\{j\kappa P_{xy,c}(t) \otimes h_{xpm,c}(t)\right\}}_{\text{Induced by Fluctuations of }c\text{-th Channel Total Power}}. \tag{6}$$

As shown in Eq. (5) and (6), the phase noise induced by SPM and XPM can be decomposed into two terms when polarization cross-terms are neglected. The second term on the right-hand side of Eq. (5) represents the SPM phase noise due to power fluctuations in the X/Y polarizations of the channel itself, while the third term accounts for the SPM phase noise arising from the total power fluctuations of the channel. A similar decomposition can be extended to the XPM phase noise induced by the $c$-th channel.

When the 4D modulation formats exhibit the constant-modulus property, the third term in the right side of Eq. (5) and (6) reduces to a constant phase rotation. And this constant phase rotation can be recovered by the carrier-phase recovery (CPR) [34]. Thus, 4D modulation formats with the constant-modulus property mitigate the phase noise induced by SPM and XPM.

## 3. Quasi-Constant modulus design

### 3.1 Principle

Although 4D modulation formats with a constant-modulus property mitigate the phase noise induced by SPM and XPM, as shown in Section II, designing nonlinearity-tolerant 4D constellations with high SE under a strict constant-modulus constraint becomes increasingly challenging due to the large number of optimization DOFs. [13]. To address this challenge, we replace the strict constant-modulus constraint with a relaxed quasi-constant modulus condition, thereby facilitating the design of QCM-QAM modulation format with relatively high SE.

QCM-QAM is constructed from the conventional PM-$M$-QAM, where $M = 2^m = |\mathcal{X}| = |\mathcal{B}|$ denotes the cardinality size of the 2D QAM constellation in each polarization, with $m$ representing the number of bits per 2D symbol. The sets $\mathcal{X}$ and $\mathcal{B}$ denote the constellation coordinates and the corresponding binary labeling, respectively. The $i$-th constellation point in $\mathcal{X}$ is denoted as $s_i = [s_{i,1}\ s_{i,2}] \in \mathbb{R}^2$ (and denoted as $s_i = [s_{i,1},\ s_{i,2},\ s_{i,3},\ s_{i,4}] \in \mathbb{R}^4$ for 4D modulation formats), and $\boldsymbol{b}_j = [b_{j,1}, b_{j,2}, \ldots b_{j,m-1}, b_{j,m}] \in \{0,1\}^m$ denotes the $j$-th binary labeling in $\mathcal{B}$, where $i=1, 2, 3, \ldots, M$ and $j=1, 2, 3, \ldots, M$.

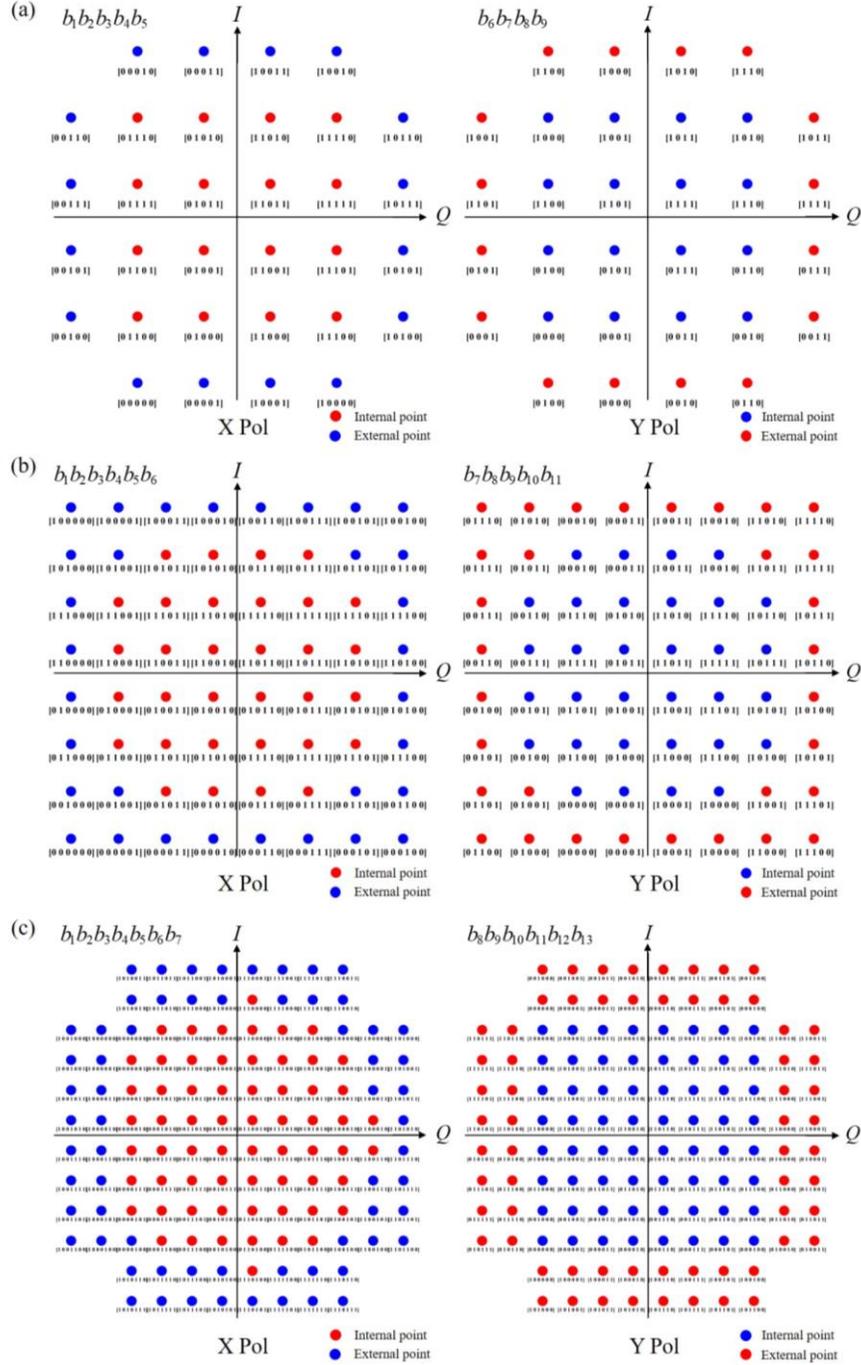

**Fig. 1.** 2D-projections of the proposed 512QCM-QAM(a), 2048QCM-QAM(b), and 8192QCM-QAM(c) modulation formats, and associated bits mapping on X/Y polarization.

To obtain the quasi-constant modulus condition in QCM-QAM, the 2D constellation points of PM-*M*-QAM for each polarization is divided into internal points (lower-energy symbols) and external points (higher-energy symbols) with the same counts, denoted as $\mathcal{X}_{inner}$ and $\mathcal{X}_{outer}$ respectively. Then the mapping of QCM-QAM, which is from the binary labeling to the 4D symbols is designed such that internal and external points never co-occur in the same 4D symbol. In other words, if one polarization carries an internal (or external) point, the other polarization must carry an external (or internal) point, thereby mitigating the total symbol energy fluctuations. Nevertheless, this design strategy reduces the achievable information rates (AIR) compared to conventional PM-M-QAM. In PM-*M*-QAM, each polarization is independently modulated with an *M*-ary QAM constellation, allowing full freedom in symbol selection across both dimensions. In contrast, QCM-QAM imposes inter-polarization constraints: the choice of a 2D symbol in one polarization restricts the candidate 2D symbols in the other to satisfy the quasi-constant modulus condition, thereby reducing the effective degrees of freedom and limiting the AIR.

Consider PM-32QAM with a SE of 10 bit/4D-sym, where the X and Y polarizations are modulated independently, each using a full 32QAM constellation with 32 freely selectable points. To construct the QCM-QAM constellation, we divide the 32QAM points in each polarization into $\mathcal{X}_{inner}$ and $\mathcal{X}_{outer}$ according to the energy level with the same counts. The 4D symbols in QCM-QAM are then formed by pairing an internal point from one polarization with an external point from the other, thereby approximating a constant total energy across the 4D space. Specifically, to obtain the quasi-constant modulus condition, only one polarization can randomly generate symbols with the origin 32QAM constellations. Once one 2D symbol in $\mathcal{X}_{inner}$ (or $\mathcal{X}_{outer}$) is generated in this polarization, only constellation points in $\mathcal{X}_{outer}$ (or $\mathcal{X}_{inner}$) can be selected to carry the information. As a result, the constellation size of the constructed QCM-QAM is $512 = 32 \times 16$, which is reduced from $1024 = 32 \times 32$ of the PM-32QAM. Consequently, each 4D symbol in the constructed QCM-QAM carries only $9 = \log_2(32 \times 16)$ bits of information, and this constructed QCM-QAM modulation formats is denoted as 512QCM-QAM. In fact, the constellation size of the QCM-QAM constructed from the PM-*M*-QAM is $M \times (M-1)$, and its SE is $(2M-1)$ bit/4D-sym. A Gray-like labeling scheme is adopted here for the QCM-QAM family, as a strict Gray mapping is difficult to achieve in 4D constellations at high SE [18], especially under quasi-constant modulus condition.

The 2D projections of the proposed 512QCM-QAM, 2048QCM-QAM and 8192QCM-QAM constellations onto the X and Y polarizations are shown in Fig. 1(a), (b) and (c), along with their corresponding binary labeling. To clearly illustrate the quasi-constant modulus property and emphasize the inter-polarization dependency, we adopt the following color-coding scheme in Fig. 3: the 2D projected constellation points in $\mathcal{X}_{inner}$ and $\mathcal{X}_{outer}$ are assigned distinct colors. A pair of 2D symbol forms a valid 4D symbol only if they share the same color, ensuring quasi-complementary energy pairing (e.g., internal–external or external–internal) across polarizations.

### 3.2 Nonlinearity tolerance of the QCM-QAM modulation format

Here, we illustrate nonlinearity tolerance of the QCM-QAM modulation format, employing the simplified NLI models has mentioned before. Eq. (5) and (6) can be rewritten as:

$$E_{x/y}^{spm}(t) \approx E_{x/y}(t) * \underbrace{\exp\left\{j\kappa |E_{x/y}(t)|^2 \otimes h_{spm}(t)\right\}}_{\text{Induced by Power Fluctuations in X/Y Pol}} * \underbrace{\exp\left\{j\kappa [P_{xy}(t) - P_{avg}] \otimes h_{spm}(t)\right\}}_{\text{Induced by Fluctuations of Total Power}} * \underbrace{\exp\left\{j\kappa P_{avg} \otimes h_{spm}(t)\right\}}_{\text{Induced by Average Total Power}}, \quad (7)$$

$$E_{x/y}^{xpm}(t) \approx E_{x/y}(t) * \underbrace{\exp\left\{j\kappa |E_{x/y,c}(t)|^2 \otimes h_{xpm,c}(t)\right\}}_{\text{Induced by Power Fluctuations in } c\text{-th Channel X/Y Pol}} * \underbrace{\exp\left\{j\kappa [P_{xy,c}(t) - P_{avg,c}] \otimes h_{xpm,c}(t)\right\}}_{\text{Induced by Fluctuations of } c\text{-th Channel Total Power}} * \underbrace{\exp\left\{j\kappa P_{avg,c} \otimes h_{xpm,c}(t)\right\}}_{\text{Induced by } c\text{-th Channel Average Total Power}}, (8)$$

where $P_{avg}$ denotes the average total power in the time domain, defined as $P_{avg} = \langle |E_x(t)|^2 + |E_y(t)|^2 \rangle$, and $\langle \cdot \rangle$ represents the expectation operator.

In Eq. (7) and (8), the second term in the right side of Eq. (5) and (6) is decomposed into two components: the second term in Eq. (7) and (8), which captures the phase noise induced by total power fluctuations, and the third term in Eq. (7) and (8), which accounts for the effect of the average total power. The second term in Eq. (7) and (8) vanishes for modulation formats with the constant-modulus property, and the quasi-constant modulus condition significantly suppresses its magnitude. Since the constant phase rotation induced by the average total power can be compensated by CPR, third term in Eq. (7) and (8) has little impact on system performance.

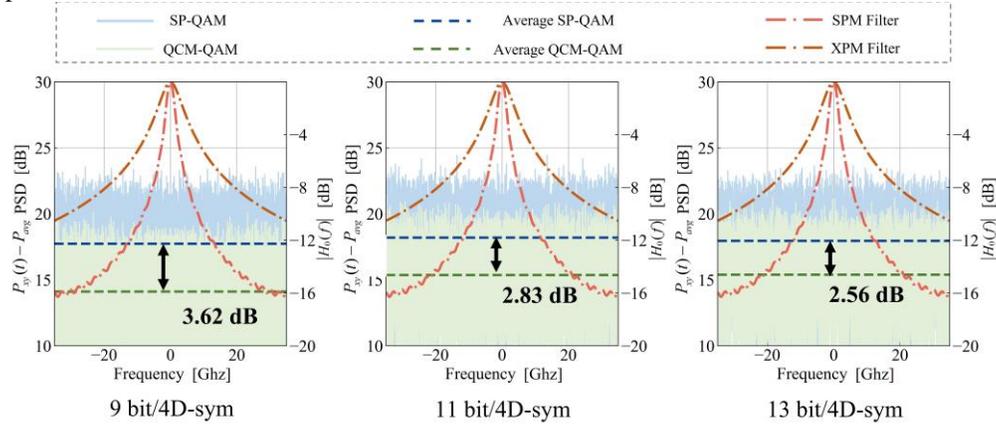

**Fig. 2.** Power spectral density (PSD) of the $P_{xy}(t) - P_{avg}$ of normalized symbol sequences that generated with the set-partitioning QAM and with the QCM-QAM, as well as the normalized magnitude frequency response of SMP filter and XPM filter of the adjacent channel.

Set-partitioning QAM [35], whose 4D modulation formats have an odd SE in bits per 4D symbol, is selected here as a benchmark for comparing nonlinearity tolerance with QCM-QAM. The PSD of $P_{xy}(t) - P_{avg}$ with the normalized symbol sequences generated using QCM-QAM and set-partitioning QAM is compared in Fig. 2, with the SE of 9, 11 and 13 bit/4D-sym. Fig. 2 also shows the normalized magnitude frequency responses of the SPM and adjacent-channel XPM filters, evaluated in an 80 km single span, 5 channel WDM system operating at 70 GBaud with 75 GHz channel spacing. The profiles of the SPM filter and the XPM filter are directly obtained with the Eq. (20) in [14] and the Eq. (8c) in [33] respectively.

As shown in Fig.2, the QCM-QAM achieves reductions of 3.62 dB, 2.83 dB, and 2.56 dB in the average PSD relative to set-partitioning QAM with the SE of 9, 11 and 13 bit/4D-sym respectively, which directly mitigates the phase noise resulted from the second term in Eq. (7) and (8). The reductions in PSD indicate that the proposed QCM-QAM modulation formats enhance tolerance to SPM and XPM, even though the strict constant modulus constraint is relaxed to a quasi-constant modulus condition. Fig. 2 shows that the reduction in PSD decreases from 3.62 dB to 2.56 dB, as the SE increases from 9 to 13 bit/4D-symbol. The reason is that the quasi-constant modulus condition more closely approximates the ideal constant modulus property at relatively low SE, where fewer constellation points exhibit distinct magnitudes.

## 4. System model and performance metrics

The unrepeatered WDM system with short transmission distances which is typical of data center interconnect (DCI) scenarios [36], is selected to investigate the nonlinear tolerance of QCM-QAM. The unrepeatered WDM system model adopted in this paper employing a bit-interleaved coded modulation (BICM) architecture, is illustrated in Fig. 3.

At the transmitter, the binary sequence is generated by a forward error correction (FEC) encoder from the input information bits. Then the bit sequence is mapped to a 4D transmitted symbol $X$ using a 4D mapper based on the designed constellation and its associated labeling pairs $\{\mathcal{X}, \mathcal{B}\}$. The 4D symbol is modulated onto two orthogonal polarizations (e.g., X and Y), and the resulting dual-polarization signal is transmitted over a single-span optical fiber link, followed by an erbium-doped fiber amplifier (EDFA). At the receiver, the received signals are processed using digital signal processing (DSP) algorithms to recover the estimated 4D symbol $Y$. Finally, $Y$ is then passed to a 4D demapper, followed by a binary FEC decoder.

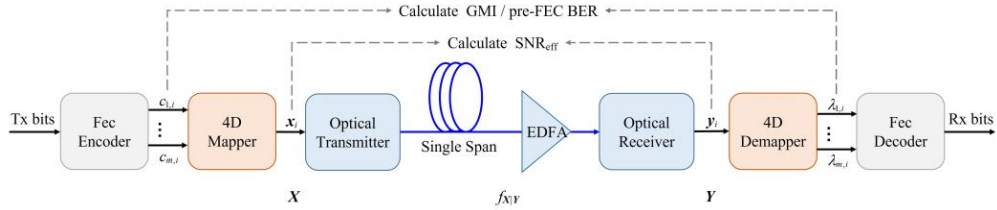

**Fig. 3.** The unrepeatered WDM system with the BICM architecture.

In the nonlinear optical fiber channel, both amplified spontaneous emission (ASE) noise and nonlinear interference degrade the SNR of the received symbols after DSP. The effective SNR (SNR$_{\text{eff}}$) is employed to represent the SNR after DSP, given by [26]

$$\text{SNR}_{\text{eff}} = \frac{P}{\sigma_{ASE}^2 + \sigma_{NLI}^2}, \quad (9)$$

where $P$, $\sigma_{ASE}^2$ and $\sigma_{NLI}^2$ represents the transmitted power, ASE noise power and the nonlinear noise power, respectively. In Eq. (9), the nonlinear noise is modeled under the Gaussian noise assumption, and its power depends on both the fiber link parameters and the modulation format, thereby reflecting the nonlinear tolerance of the latter.

The AIR is defined as the amount of information per symbol that can be reliably transmitted over a given channel [37]. The AIR over a given channel with different coded modulation schemes can be quantified by mutual information (MI) or generalized mutual information (GMI). In general, MI applies to nonbinary codes or multilevel coded modulation with multistage decoding, whereas GMI is more appropriate for simpler BICM systems. As a robust performance metric for optical system with BICM scheme [38], GMI can be estimated via Gauss-Hermite quadrature (GHQ) and is given by Eq. (17) and (18) in [37]

$$\text{GMI} = G(\mathcal{X}, \mathcal{B}, f_{X|Y}) = m + \frac{1}{M} \sum_{k=1}^{m} \sum_{b \in (0,1)} \sum_{i \in \mathcal{I}_k^b} \int_{\mathcal{R}^N} f_{X|Y}(\boldsymbol{y}|\boldsymbol{x}_i) \cdot \log \frac{\sum_{j \in \mathcal{I}_k^b} f_{X|Y}(\boldsymbol{y}|\boldsymbol{x}_j)}{\frac{1}{2}\sum_{p=1}^{M} f_{X|Y}(\boldsymbol{y}|\boldsymbol{x}_p)} d\boldsymbol{y}, \quad (10)$$

where $\mathcal{I}_k^b$ is the set of constellation points associating the $k$-th bit to the bit value $b \in \{0,1\}^m$. A mismatched receiver is employed in the GMI calculation here. Specifically, the channel law is approximated by a Gaussian distribution, which is a common approach for estimating GMI in fiber optic systems.

## 5. Numerical results and analysis

In this section, QCM-QAM modulation formats are compared with set-partitioned QAM [35], which achieves an odd SE in bits per 4D symbol. Both modulation schemes are implemented in an system over short transmission distances, as illustrated in Fig. 3.

The unrepeatered WDM system employs five co-propagating channels, each operating at a symbol rate of 70 GBaud, with a WDM channel spacing of 75 GHz and a root-raised-cosine (RRC) pulse shaping filter with a roll-off factor of 0.05. Two types of optical fiber are considered here: standard single-mode fiber (SSMF) and non-zero dispersion-shifted fiber (NZDSF), to evaluate the nonlinear performance of QCM-QAM. The considered simulation parameters of the transmission system are shown in Table 1. At the receiver, ideal dispersion compensation is applied, followed by matched filtering and downsampling. Finally, the average phase rotation is compensated.

**Table 1. Simulation Parameters**

| Parameter Name | SSMF | NZDSF |
| --- | --- | --- |
| Symbol rate (GBaud) | 70 | 70 |
| Root-raised-cosine roll-off factor | 0.05 | 0.05 |
| Center wavelength (nm) | 1550 | 1550 |
| Attenuation (dB/km) | 0.21 | 0.2 |
| Dispersion parameter (ps nm$^{-1}$ km$^{-1}$) | 16.9 | 3.9 |
| Nonlinearity parameter (W$^{-1}$ km$^{-1}$) | 1.31 | 1.6 |
| Fiber span length (km) | 80 | 80 |
| EDFA noise figure (dB) | 4.5 | 4.5 |

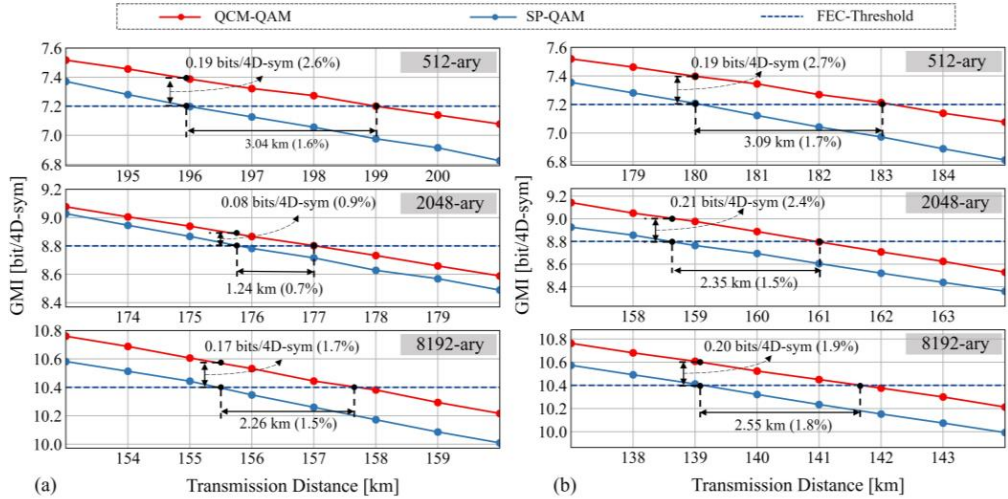

**Fig. 4.** GMI versus transmission reach for 512-, 2048-, and 8192-ary modulation formats at optimal launch power over (a) SSMF and (b) NZDSF.

The 512-, 2048-, and 8192-ary modulation formats, corresponding to SE of 9, 11, and 13 bit/4D-sym, are compared near the threshold of soft-decision forward error correction (SD-FEC) based on BICM [18] with a code rate of 0.8. Fig. 4(a) and (b) show GMI as a function of transmission distance over SSMF and NZDSF, respectively. QCM-QAM demonstrates improved transmission reach compared to SP-QAM in both fiber types. Over SSMF, it achieves reach gains of 1.6%, 0.9%, and 1.7% for 512-, 2048-, and 8192-ary modulation formats (corresponding to spectral efficiencies of 9, 11, and 13 bit/4D-sym), respectively. At the soft-decision forward error correction (SD-FEC) threshold (code rate = 0.8), QCM-QAM also provides GMI gains of 0.19, 0.08, and 0.17 bit/4D-sym at these SEs. In NZDSF, the higher reach improvement is obtained. QCM-QAM extends transmission distance by 1.7%, 1.5%, and 1.8% over SP-QAM for the same modulation formats. Correspondingly, GMI gains of 0.19, 0.21, and 0.20 bit/4D-sym are observed at the SD-FEC threshold. The greater improvement stems from the enhanced nonlinear tolerance afforded by QCM-QAM. Because signals experience stronger NLI in NZDSF, the quasi-constant modulus property of QCM-QAM yields more pronounced gains in both transmission reach and GMI compared to SSMF.

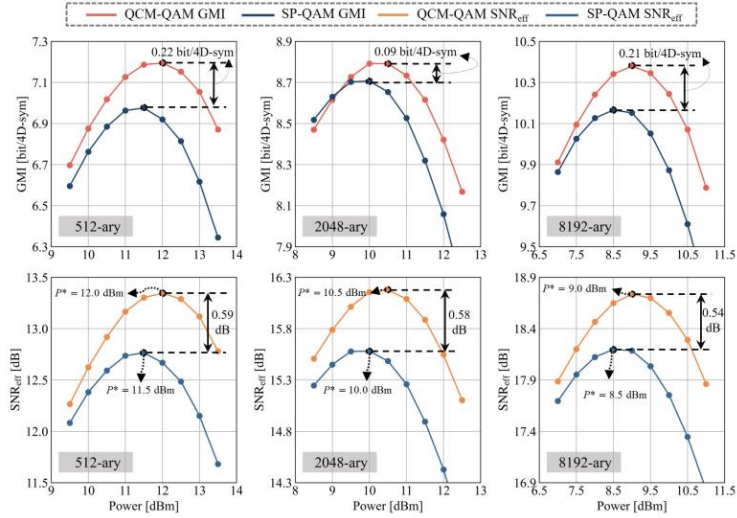

**Fig. 5.** GMI (top) and effective SNR (bottom) versus launch power for 512-, 2048-, and 8192-ary modulation formats after transmission over 199 km, 177 km, and 158 km of SSMF, respectively.

Fig. 5 shows the GMI and effective SNR as functions of launch power in SSMF for QCM-QAM and SP-QAM in the top and bottom parts, respectively. Results are presented for 512-, 2048-, and 8192-ary modulation formats over transmission links of 199 km, 177 km, and 158 km, respectively. For all SEs, QCM-QAM exhibits a clear increase in optimal launch power compared to SP-QAM, as observed in both the GMI and effective SNR curves. This shift toward higher optimal power indicates that QCM-QAM exhibits superior nonlinear tolerance contrasted with SP-QAM.

The GMI estimated in SSMF transmission for QCM-QAM and SP-QAM is shown in the three subfigures in the top part of Fig. 5. In the linear regime, i.e., at low launch powers, the performance gain offered by QCM-QAM is minimal across all SEs. In fact, a slight GMI degradation is observed for 2048QCM-QAM compared to 2048SP-QAM. This stems from the reduced linear performance of QCM-QAM, which arises from the Gray-like labeling scheme adopted to preserve its quasi-constant modulus property. However, at the optimal launch power, QCM-QAM demonstrates clear advantages due to its enhanced nonlinear tolerance, achieving GMI gains of 0.22, 0.09, and 0.21 bit/4D-sym for the 512-, 2048-, and 8192-ary formats, respectively.

The effective SNR estimated in SSMF transmission for QCM-QAM and SP-QAM is shown in the bottom part of Fig. 5. As launch power increases, the effective SNR gain provided by

QCM-QAM also increases, reflecting its enhanced mitigation of NLI, which scales approximately with the cube of the launch power [26]. At the optimal launch power, QCM-QAM achieves effective SNR gains of 0.59 dB, 0.58 dB, and 0.54 dB for the 512-, 2048-, and 8192-ary modulation formats, respectively.

Fig. 6 illustrates the GMI and effective SNR as functions of launch power in NZDSF for both QCM-QAM and SP-QAM. The top and bottom parts show the GMI and effective SNR, respectively, for 512-, 2048-, and 8192-ary modulation formats over transmission distances of 183 km, 161 km, and 142 km. As observed in SSMF transmission, QCM-QAM consistently shifts the optimal launch power to higher values across all SEs in both GMI and effective SNR curves, reflecting its enhanced tolerance to fiber nonlinearities compared to SP-QAM.

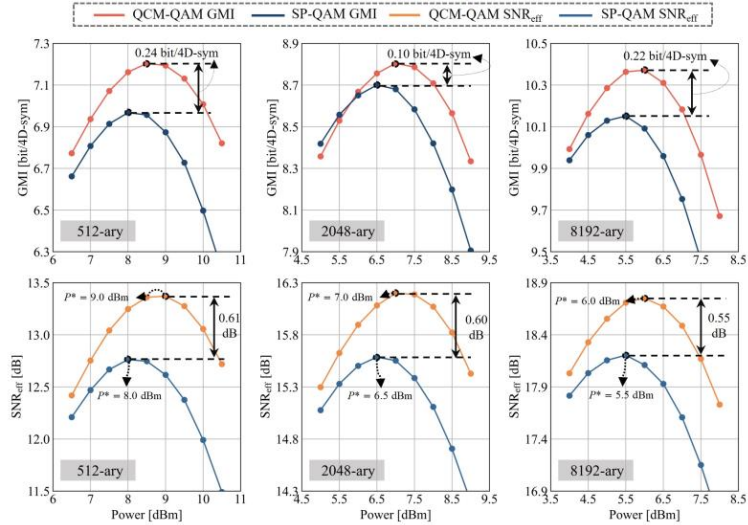

**Fig. 6.** GMI (top) and effective SNR (bottom) versus launch power for 512-, 2048-, and 8192-ary modulation formats after transmission over 142 km, 161 km, and 183 km of NZDSF, respectively.

The top part of Fig. 6 shows the estimated GMI for QCM-QAM and SP-QAM in NZDSF transmission. Similar trends are observed as in SSMF: QCM-QAM provides minimal gain in the linear regime (low launch power), but demonstrates a clear advantage at the optimal launch power. Specifically, GMI gains of 0.24, 0.10, and 0.22 bit/4D-sym are achieved for the 512-, 2048-, and 8192-ary formats, respectively, exceeding the corresponding gains in SSMF. This enhancement stems from QCM-QAM's superior nonlinear tolerance. Because NZDSF exhibits stronger nonlinear impairments than SSMF, the quasi-constant modulus property of QCM-QAM yields greater performance benefits in this fiber type.

The bottom part of Fig. 5 shows the estimated effective SNR for QCM-QAM and SP-QAM in NZDSF transmission. Similarly, the effective SNR gain provided by QCM-QAM increases as the launch power increases. Same with the estimated GMI, higher gains of effective SNR are achieved in NZDSF. Specifically, gains of 0.61, 0.60, and 0.55 dB are achieved for the 512-, 2048-, and 8192-ary formats.

Regardless of whether transmission occurs over SSMF or NZDSF, the effective SNR gain decreases with increasing SE. Specifically, in SSMF transmission, the effective SNR gain decreases from 0.59 dB to 0.54 dB, while in NZDSF transmission, it decreases from 0.61 dB to 0.55 dB as SE increases from 9 bit/4D-sym to 13 bit/4D-sym. This observation aligns with the analysis of nonlinear tolerance discussed in Section III. The nonlinear tolerance of QCM-QAM is attributed to its quasi-constant modulus property, which suppresses the magnitude of $P_{xy}(t) - P_{avg}$ to mitigate NLI. As the modulation format's SE increases, extending the 4D constellation space for shaping, the suppression of $P_{xy}(t)-P_{avg}$ becomes more challenging. As shown in Fig. 2, the reduction in average PSD of $P_{xy}(t)-P_{avg}$ achieved by QCM-QAM decreases from 3.62 dB to 2.56 dB as the SE increases from 9 bit/4D-sym to 13 bit/4D-sym. The decrease in the PSD reduction means the higher NLI power as shown in Eq. (7) and (8), which in turn reduces the effective SNR gain.

Fig. 7 shows normalized 2D projections of the received 4D symbols after Rx DSP. Without loss of generality, the 512-ary modulation format is selected to compare QCM-QAM and SP-QAM. Fig. 7(a) displays the received symbols after transmission over 160 km of SSMF at a transmitted power of 12 dBm, while Fig. 7(b) shows those after transmission over 145 km of NZDSF at 9 dBm. It is observed that QCM-QAM reduces nonlinear phase noise in both transmission scenarios, particularly for symbols with high energy, which are more severely affected by Kerr nonlinearity.

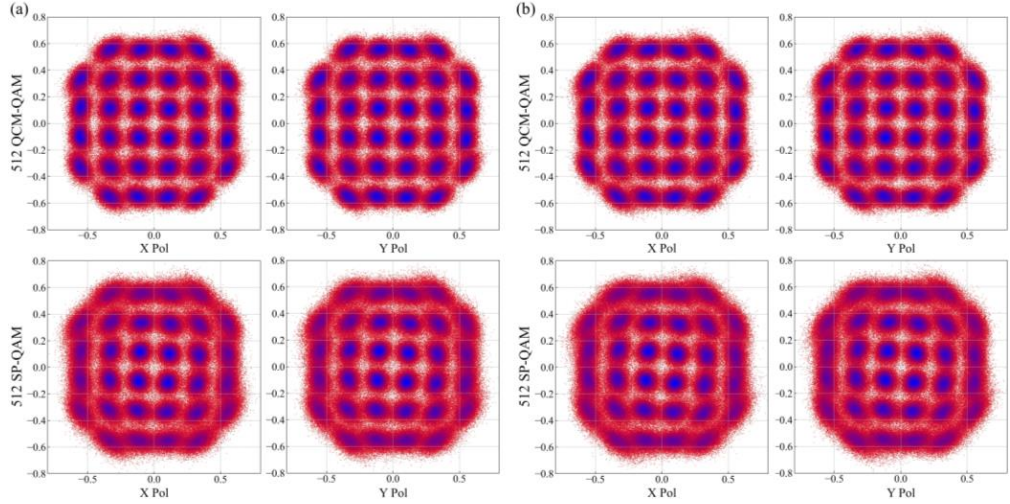

**Fig. 7.** Normalized 2D projections of the received 4D symbols after Rx DSP. (a) Symbols after transmission over 160 km of SSMF at a transmitted power of 12 dBm. (b) Symbols after transmission over 145 km of NZDSF at a transmitted power 9 dBm.

## 6. Conclusion

In this paper, we analyze the quasi-constant modulus property and harness it to design nonlinearity-tolerant 4D modulation formats. Specifically, we propose a family of QCM-based QCM-QAM constellations with high SEs of 9, 11, and 13 bit/4D-sym, corresponding to 512-ary, 2048-ary, and 8192-ary 4D constellations, respectively. By maintaining a quasi-constant signal energy, QCM-QAM provides comparable mitigation of SPM and XPM, as confirmed by both theoretical analysis and numerical simulations. The nonlinear tolerance of QCM-QAM across different SEs is validated in unrepeatered WDM transmission over both SSMF and NZDSF, where a consistent shift of the optimal launch power to higher values and a considerable improvement in effective SNR are observed. Despite the slight linear performance penalty introduced by the Gray-like labeling in QCM-QAM, GMI gains are still realized due to its strong nonlinear tolerance.

In SSMF, QCM-QAM achieves effective SNR gains of 0.59 dB, 0.58 dB, and 0.54 dB, and GMI gains of 0.22, 0.09, and 0.21 bit/4D-sym, after transmission over 199 km, 177 km, and 158 km, respectively. In NZDSF, it yields effective SNR gains of 0.61 dB, 0.60 dB, and 0.55 dB, and GMI gains of 0.24, 0.10, and 0.22 bit/4D-sym, over 183 km, 161 km, and 142 km, respectively. These results correspond to modulation formats with SEs of 9, 11, and 13 bit/4D-sym, demonstrating that QCM-QAM provides robust nonlinear mitigation across both fiber types and a wide range of SEs. Furthermore, QCM-QAM achieves reach extensions of 1.6%, 0.9%, and 1.7% in SSMF, and 1.7%, 1.5%, and 1.8% in NZDSF, for the 512-ary, 2048-ary, and 8192-ary modulation formats, respectively.

We believe that QCM-QAM is a promising candidate for transmission systems operating in highly nonlinear regimes. In the future, the binary switching optimization (BSO) algorithm [18, 39] could be employed to improve labeling by adaptively switching the bit-to-symbol mappings. Furthermore, the quasi-constant modulus property could be incorporated as a constraint in constellation optimization for enhanced nonlinear tolerance, enabling joint optimization of both the binary labeling and the symbol coordinates.

**Funding.** National Key R&D Program of China (2023YFB2905400); National Natural Science Foundation of China (62025503); Shanghai Jiao Tong University 2030 Initiative.

**Disclosures.** The authors declare no conflicts of interest.

**Data availability.** Data underlying the results presented in this paper are not publicly available at this time but maybe obtained from the authors upon reasonable request.